\documentclass[aps,prc,twocolumn,times,graphicx,tighten,showpacs]{revtex4}
\usepackage[dvips]{graphicx}

\newcommand{\bea}{\begin{eqnarray}}
\newcommand{\eea}{\end{eqnarray}}
\newcommand{\be}{\begin{equation}}
\newcommand{\ee}{\end{equation}}

\begin{document}

\title{
Superscaling in charged current neutrino quasielastic scattering in the
relativistic impulse approximation}

\author{J.A. Caballero$^a$, J.E. Amaro$^b$, M.B. Barbaro$^c$,
T.W. Donnelly$^d$, C. Maieron$^e$ and J.M. Udias$^f$}

\affiliation{$^a$Departamento de F\'{\i}sica At\'omica, Molecular y Nuclear,
Universidad de Sevilla, 41080 Sevilla, SPAIN}
\affiliation{$^b$Departamento de F\'{\i}sica Moderna, Universidad de Granada,
  18071 Granada, SPAIN}
\affiliation{$^c$Dipartimento di Fisica Teorica, Universit\`a di Torino and
  INFN, Sezione di Torino, Via P. Giuria 1, 10125 Torino, ITALY}
\affiliation{$^d$Center for Theoretical Physics, Laboratory for Nuclear
  Science and Department of Physics, Massachusetts Institute of Technology,
  Cambridge, MA 02139, USA}
\affiliation{$^e$INFN, Sezione di Catania, Via Santa Sofia 64, 95123 Catania,
  ITALY}
\affiliation{$^f$Departamento de F\'{\i}sica At\'omica, Molecular y Nuclear,
Universidad Complutense de Madrid, 28040 Madrid, SPAIN}



\begin{abstract}
  Superscaling of the
  quasielastic cross section in charged current neutrino-nucleus reactions at
  energies of a few GeV is investigated within the framework of the
  relativistic impulse approximation.  Several approaches are used to describe
  final state interactions and comparisons are made with the plane wave
  approximation.  Superscaling is very successful in all
  cases.  The scaling function obtained using a relativistic mean field for the final states
  shows an asymmetric shape with a long tail extending towards
  positive values of the scaling variable, in excellent agreement with the behavior
  presented by the experimental scaling function.
\end{abstract}


\maketitle

In the context of inclusive quasielastic (QE) electron scattering
at intermediate to high energies, the concepts of
scaling~\cite{West74} and superscaling~\cite{DS199} have been explored in
previous work~\cite{DS299,MDS02}, where an exhaustive analysis of
the $(e,e')$ world data demonstrated the quality of the scaling
behavior. Scaling of the first kind (no dependence
on the momentum transfer) is reasonably well respected at
excitation energies below the QE peak, whereas scaling of second
kind (no dependence on the nuclear species) is excellent in the
same region. The simultaneous occurrence of both kinds of scaling
is called superscaling. At energies above the QE peak both scaling
of the first and, to a lesser extent, of the second kind are shown
to be violated because of important contributions introduced by
effects beyond the impulse approximation, namely, inelastic
scattering~\cite{Alvarez-Ruso:2003gj} together with correlations
and meson exchange currents in both the 1p-1h and 2p-2h
sectors~\cite{Amaro:2001xz,DePace}.

The scaling analysis of $(e,e')$ data has recently been extended
through the QE peak into the $\Delta$ region~\cite{neutrino1}. Of
relevance to the present work we note that the high-energy
inclusive electron scattering cross section is well represented up
to the $\Delta$ peak using the scaling ideas, importantly, with an
{\sl asymmetric} QE scaling function. In that study the scaling
approach was also used to predict nuclear $(\nu,\mu)$ cross
sections, based on the assumption of a universal scaling function,
valid for both electron and neutrino scattering
at
corresponding kinematics.

In this letter we investigate the QE scaling properties of
charged-current (CC) neutrino-nucleus scattering within the
context of the Relativistic Impulse Approximation (RIA).  After
verifying that various RIA models do superscale, we compare the
associated scaling functions with the $(e,e')$ phenomenological
one referred to above. This allows a check on the consistency of
the universality assumption and on the capabilities of different
models to yield
the required properties of the experimental
scaling function, specifically, its asymmetric form.

Here we follow the general procedure of scaling and superscaling
studies, namely we first construct inclusive cross sections within
a model and then obtain scaling
functions by dividing them by the relevant single-nucleon cross
sections weighted by the corresponding proton and
neutron numbers~\cite{Barbaro:1998gu,DS299,MDS02}. The scaling function is
plotted against the scaling variable $\psi(q,\omega)$, with $q$
and $\omega$ the momentum and energy transferred in the process,
and its scaling properties analyzed.

Within the RIA framework CC neutrino-nucleus QE scattering is
described by assuming that at the $\nu$-$\mu$ vertex one vector
boson is exchanged with the nucleus, interacting with only one
nucleon, which is then emitted while the remaining (A-1) nucleons
in the target remain as spectators. The nuclear current operator
is thus taken to be the sum of single-nucleon currents, for which
we employ the usual relativistic free nucleon
expressions~\cite{Alb97,neutrino1}. The RIA approach has been
extensively and successfully applied in investigations of
exclusive electron scattering reactions~\cite{Udias}. Further
details on the model have been presented in
\cite{Alb97,Alb98,Chiara03}.

We describe the bound nucleon states as self-consistent
Dirac-Hartree solutions, derived within a Relativistic Mean Field
(RMF) approach using a Lagrangian containing $\sigma$, $\omega$
and $\rho$ mesons~\cite{boundwf}. For the description of the
outgoing nucleon states we consider several different approaches.
In one we use plane-wave spinors (thus no Final-State Interactions
(FSI)), corresponding to the Relativistic Plane-Wave Impulse
Approximation (RPWIA). However, comparisons with data require a
more realistic description of the final nucleon state, which
should include the effects due to FSI. This is accomplished by
using solutions of a Dirac equation
containing relativistic potentials. This constitutes  the Relativistic
Distorted-Wave Impulse Approximation (RDWIA).
%
%

%
%
The use of complex relativistic optical potentials fitted to
elastic proton scattering data has proven to be successful in
describing {\em exclusive} $(e,e'p)$ scattering
reactions~\cite{Udias}. The absorption produced by the imaginary
term represents the loss of flux into inelastic channels. For {\em
inclusive} processes such as $(e,e')$ and $(\nu,\mu)$, where a
selection of the exclusive single-nucleon knockout channel cannot
be made, the contribution from these inelastic channels should be
retained. Ignoring them would lead to an underestimation of the
inclusive cross section~\cite{Chiara03}. A simple way of obtaining
the right inclusive strength within the RIA is to use purely real
potentials.
We consider two choices for the real part. The first
uses the phenomenological relativistic optical potential from the
energy-dependent, $A$-independent parameterizations (EDAIC, EDAIO,
EDAICa) derived by Clark {\it et al.} \cite{Clark}, but with their
imaginary parts set to zero. The second approach employs distorted
waves obtained with the same relativistic mean field used to
describe the initial bound nucleon states. We refer to these two
FSI descriptions as real Relativistic Optical Potential (rROP) and
Relativistic Mean Field (RMF), respectively.
%
%
%
%
Dispersion relation and Green function techniques have been used
to derive rigorously the potentials for inclusive scattering (see
refs.~\cite{Horikawa1980,Chinn89,Giusti2003}) leading to results
which are within a few percent of those obtained in the IA with
either the rROP~\cite{Giusti2003} or the mean
field~\cite{Horikawa1980}.
Note that
the RMF model (1) is known to work quite well for inclusive QE
$(e,e')$ scattering (which is verified here; see below) and (2) is
constructed in a way that fulfills the dispersion
relation~\cite{Horikawa1980} and maintains the continuity equation.

\begin{figure}[tph]
\begin{center}
\includegraphics[scale=0.5,  bb= 80 500 500 750]{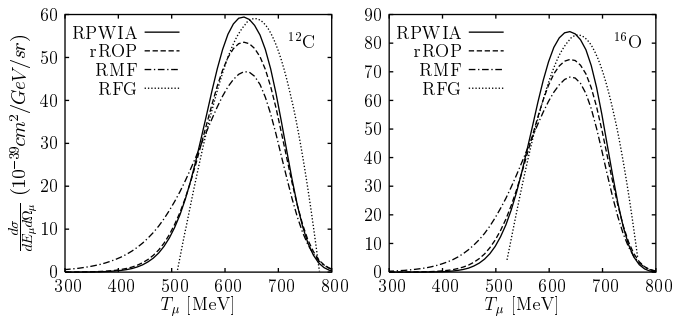}
\caption{
  Quasielastic differential cross section $d\sigma/dE_\mu d\Omega_\mu$
  versus the muon kinetic energy $T_\mu$ for the reaction
  $(\nu_\mu,\mu^-)$ on $^{12}$C (left) and $^{16}$O (right). The incident
  neutrino energy is $\varepsilon_\nu=1$ GeV and the muon
  scattering angle is $\theta_\mu=45^o$. In each panel we present
  results for RPWIA (solid), rROP (dashed) and RMF (dot-dashed). The cross
  section for the RFG model is also presented for reference (dotted).}
\label{Fig1}
\end{center}
\end{figure}

Our results for CC neutrino-nucleus QE scattering are presented in
Fig.~\ref{Fig1}, where we show the differential cross section
$(d\sigma/dE_\mu d\Omega_\mu)$ as a function of the outgoing muon
kinetic energy for $^{12}$C and $^{16}$O.  For reference we also
include the results obtained within the Relativistic Fermi Gas
(RFG) (dotted) with Fermi momenta $k_F=228$ MeV/c for
$^{12}$C~\cite{MDS02} and $k_F=216$ MeV/c for $^{16}$O~\cite{Co}.
The RFG curves contain a phenomenological energy
shift~\cite{MDS02} yielding RFG cross sections which are very
similar to the RPWIA results.

The mean field dynamics in the initial and final nuclear states
lead to cross sections having tails that extend both below and
above the kinematical region where the RFG is defined. With FSI
included we observe a reduction of the cross section, particularly
in the case of the RMF potential where it is seen to be about
$20\%$ for both nuclei in the region close to the maximum. Notice
also a slight displacement in the maximum of the cross section in
the cases of the two RIA-FSI models. However, the most striking
feature is the long tail displayed by the RMF cross section for
small muon kinetic energies.  This corresponds to transferred
energies above the QE peak, i.e., positive values of the scaling
variable. These FSI effects lead to a clear asymmetry in the RMF
cross section,  in contrast to the RPWIA and rROP results. The
discrepancy between the two FSI approaches is linked to the
different behaviors of the two potentials: for high nucleon
kinetic energies (small muon energies) the scalar and vector
energy-dependent potentials of the rROP model are significantly
reduced with respect to the RMF~\cite{Udias}. The latter approach
has strong scalar and vector potentials that are delicately balanced 
%
%
%
%
and that shift the strength towards higher energies.
This explains why the rROP cross section is closer to the RPWIA
case, and moreover, why the main difference between the two models
is observed in the region of lower muon energies. Note that the
asymmetric broadening of the RMF cross section is similar to that
observed in non-relativistic models of the FSI, described by some
as medium modifications of the p-h propagator~\cite{Co}.

Let us now study the superscaling properties. We present results
for the scaling function $f(\psi^\prime)$, which is obtained by
dividing the calculated differential cross section of
Fig.~\ref{Fig1} by the single-nucleon cross section as given in
Eqs.~(45,52,86-94) of \cite{neutrino1}. This function is plotted
against the shifted QE scaling variable $\psi^\prime$ defined as
\begin{equation}
\psi^\prime\equiv\frac{1}{\sqrt{\xi_F}}\frac{\lambda^\prime-\tau^\prime}
 {\sqrt{(1+\lambda^\prime)\tau^\prime+
\kappa\sqrt{\tau^\prime(1+\tau^\prime)}}} \, ,
\label{eq1}
\end{equation}
where $\lambda^\prime\equiv (\omega-E_{shift})/2m_N$,
$\kappa\equiv q/2m_N$, $\tau^\prime\equiv \kappa^2-\lambda^{\prime
2}$, and $\xi_F\equiv\sqrt{1+(k_F/m_N)^2}-1$. The energy shift
$E_{shift}$ has been taken from~\cite{MDS02}. Results correspond
to fixed scattering muon angle $\theta_\mu=45^o$, although similar
scaling functions are obtained with other values.

\begin{figure}[tph]
\begin{center}
\includegraphics[scale=0.52,  bb= 150 190 420 740]{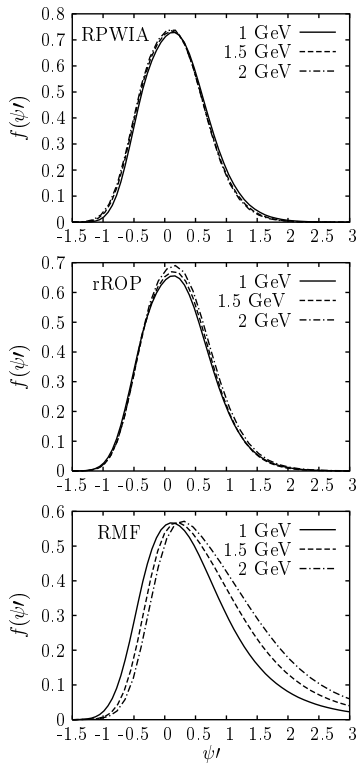}
\caption{Scaling function for three values of the incident neutrino energy
  $\varepsilon_\nu$.  Results correspond to $^{12}$C and $\theta_\mu=45^o$.
  Top, middle and bottom panels refer to RPWIA, rROP and RMF models (see
  text for details).}
\label{Fig2}
\end{center}
\end{figure}

\begin{figure}[tph]
\begin{center}
\includegraphics[scale=0.52,  bb= 150 190 420 740]{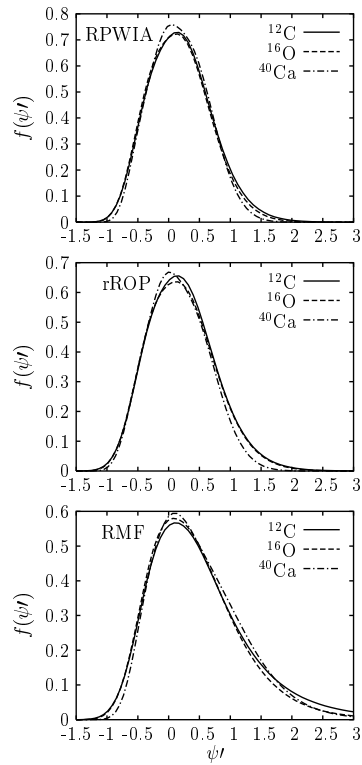}
\caption{Scaling function for three nuclei: $^{12}$C, $^{16}$O and $^{40}$Ca.
  Results correspond to $\varepsilon_\nu=1$ GeV and $\theta_\mu=45^o$.  Top,
  middle and bottom panels as in previous figure.}
\label{Fig3}
\end{center}
\end{figure}

Scaling of the first kind is explored in Fig.~\ref{Fig2}, where we
present $f(\psi^\prime)$ for $^{12}C$ at three different values of
the incident energy. Each panel in the figure corresponds to a
different description of the FSI. As one can see, the scaling
function for the RPWIA and rROP models shows a very mild
dependence on the momentum transfer in both positive and negative
$\psi'$ regions. In the case of the RMF, a slight shift occurs in
the so-called ``scaling region'' $\psi^\prime<0$, whereas for
$\psi^\prime$ positive the model breaks scaling at roughly the
30\% level in the energy region explored here. This is not in
conflict with the experimental $(e,e')$ data that indeed leave
room for some breaking of first-kind scaling in this region, due
partly to $\Delta$ production and partly to other contributions,
such as MEC and their associated correlations in the 2p-2h
sector~\cite{DePace}. It is striking that the RMF model, in spite
of being based on the impulse approximation, leads to the same
kind of behavior which is apparently not reproduced by
uncorrelated models in the impulse approximation. Importantly, the
RMF scaling function exhibits a significant asymmetry, being
larger for positive $\psi^\prime$, which persists for all neutrino
energies (actually increasing with $\varepsilon_\nu$).

Scaling of the second kind is studied in Fig.~\ref{Fig3}, where
the scaling function evaluated for three nuclei is presented using
the three models. The values of the Fermi momentum used range from
$k_F=216$ MeV/c for $^{16}$O to $k_F=241$ MeV/c for
$^{40}$Ca~\cite{MDS02}. The initial bound states have been
obtained using the parameters of the set NLSH~\cite{Sharma}.
Results with other parameterizations are similar and do not change
the general conclusions. As observed, the differences introduced
by changing nucleus are small. We may conclude that within the
present model scaling of second kind is very successful. Indeed,
this is just what is seen experimentally, at least for
$\psi^\prime < 0$ where scaling of the second kind is
excellent~\cite{DS199,DS299}.

\begin{figure}[h]
\begin{center}
\includegraphics[scale=0.51,  bb= 140 380 440 750]{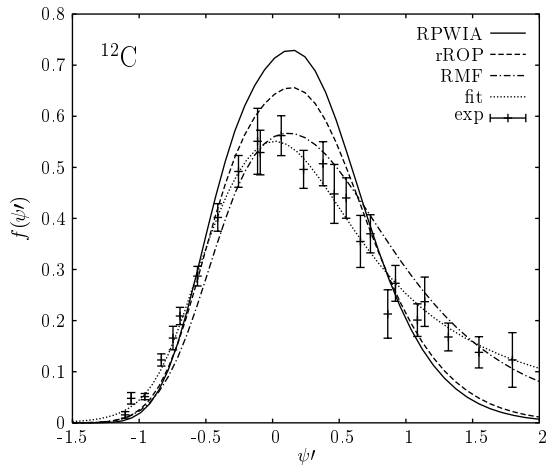}
\caption{Scaling function evaluated within the RPWIA (solid), rROP
(dashed) and RMF (dot-dashed) approaches compared with the
averaged experimental function together with a phenomenological
parameterization of the data (dotted).} \label{Fig4}
\end{center}
\end{figure}

Finally, in Fig.~\ref{Fig4} we compare the model superscaling
functions with the averaged QE phenomenological function obtained
from the analysis of $(e,e')$ data~\cite{DS199,neutrino1,Jourdan}.

First we observe the symmetric character of the RPWIA and rROP
results, which clearly differ from the experimental function. On
the contrary, the RMF curve displays a pronounced tail that
extends toward positive values of $\psi^\prime$, following closely
the asymmetric behavior of the data and yielding excellent
agreement with the phenomenological scaling function. In the light
of the analysis in~\cite{neutrino1} this immediately implies that
the RMF approach yields excellent agreement with the experimental
$(e,e')$ inclusive cross section. The asymmetric shape of the RMF
result constitutes a basic difference not only from the other two
models explored in this work, but also from other modeling
presented in the literature, such as those in~\cite{neutrino2},
where the long tail in the superscaling function is absent.
Indeed, the CDFM model for
correlation effects presented in~\cite{neutrino2} is
manifestly symmetrical around the
QE peak. It should be remarked that all of the curves in Fig.~\ref{Fig4}
essentially satisfy the Coulomb sum rule, i.e., they integrate to
unity.

The results in Fig.~\ref{Fig4} 
demonstrate that the RMF model
is successful at incorporating important dynamical effects
missed by the other models~\cite{highp}. Only in the RMF case is
the required asymmetric form of the scaling function obtained. It
should be stressed that the main factors responsible for the
asymmetry in the calculated scaling function are not only
relativity (present in all models considered here), but the
particular description of the final continuum nucleon states. The
asymmetry observed in the data has usually been ascribed to the
role played by ingredients beyond the mean field, such as
short-range correlations and two-body currents~\cite{DePace}.
%
%
%
%
We show here that it can be explained within the RIA framework
resorting only to one-body excitations, provided strong
relativistic potentials are included in the model. It would be
interesting to see how a more elaborated formalism that includes
multinucleon excitations as in~\cite{Chinn89} compares with
superscaling data.

In summary, we have shown that superscaling is fulfilled to high
accuracy within the present relativistic impulse approximation in
the QE region, and that this holds for the three different
descriptions of FSI considered here. The asymmetric shape and the
long tail at positive $\psi^\prime$-values observed in the
experimental scaling function is reproduced only by the RMF model.
This result 
reinforces our confidence in the adequacy of
descriptions of FSI effects for inclusive $(e,e')$ and $(\nu,\mu)$
reactions when based on the RMF approach.

\section*{Acknowledgements}
This work was partially supported by DGI (Spain): BFM2002-03315,
BFM2002-03218, FPA2002-04181-C04-04, BFM2003-041-C02-01, by the
Junta de Andaluc\'{\i}a, and by the INFN-CICYT  collaboration
agreement N$^o$ 04-17. It was also supported in part (TWD) by U.S.
Department of Energy under cooperative agreement No.
DE-FC02-94ER40818.


\end{document}